\documentclass[aps,eqsecnum,nofootinbib,showpacs]{revtex4}
\usepackage[dvips]{graphicx}  % for graphics
\usepackage{amsmath,amsfonts}
\usepackage{bm}
\usepackage{color}  % for color
\begin{document}
%%%%%%%%%%%%%%%%  Title page  %%%%%%%%%%%%%%%%%%%%
%%
\title{Gravity in a warped 6D world with an extra 2D sphere}
\author{Akira Kokado}
\email{kokado@kobe-kiu.ac.jp}
\affiliation{Department of Physical Therapy, Kobe International University, Kobe 658-0032, Japan}
\author{Takesi Saito}
\email{tsaito@k7.dion.ne.jp}
\affiliation{Department of Physics, Kwansei Gakuin University,
Sanda 669-1337, Japan}
\date{\today}
\begin{abstract}
Corrections to Newton's inverse law have been so far considered, but not clear in warped higher dimensional worlds, because of complexity of the Einstein equation. Here we give a model of a warped 6D world with an extra 2D sphere. We take a general energy-momentum tensor, which does not depend on a special choice of bulk matter fields. The 6D Einstein equation reduces to the spheroidal differential equation, which can be easily solved. The gravitational potential in our 4D universe is calculated to be composed of infinite series of massive Yukawa potentials coming from the KK mode, together with Newton's inverse law. The series of Yukawa type potentials converges well to behave as $1/r^3$  near $r=0$.
\end{abstract}

\maketitle
%%%%%%%%%%%%%%%%%%%%%%%%%%%%%%%%%%%%%%%%%%%%%%%%%%
\section{Introduction}\label{sec:intro}
%%%%%%%%%%%%%%%%%%%%%%%%%%%%%%%%%%%%%%%%%%%%%%%%%%
%
%
There is a long history in the search of corrections to Newton's inverse law in gravity. Especially, gravity in higher dimensional worlds has been drawn many interests, because the gravitational potential is expected to reflect sharply on number of dimensions. This kind of work, however, turns out to be extremely hard, because of complexity of the Einstein equation in higher dimensions. So, this problem of corrections to Newton's inverse law has been still not clear, especially in warped higher dimensional worlds.  \\
\indent In this paper we concentrate on gravity in the 6D world with an extra 2D sphere. Here, the extra 2D sphere preserves spherically symmetric, whereas the 4D world carries metrics with a warp factor $\phi (\theta )$. The compact 2D extra space is expected to induce the gravitational potentials of Yukawa type coming from KK modes, which are corrections to Newton's inverse law. The Yukawa potential appears in several theoretical models, such as unification theories that predict new fundamental interactions with a massive gauge boson, a massive Brans-Dicke scalar, a light dilaton and so on \cite{ref:Fujii, ref:Bars, ref:Donoghue, ref:Floratos, ref:Kehagias, ref:Kaplan, ref:Hees}. Randall-Sundrum summed up the series of Yukawa potentials in the warped 5D world with branes  \cite{ref:Randall}. Salvio has discussed the same potential in the 6D supergravity \cite{ref:Salvio, ref:Parameswaran1, ref:Parameswaran2, ref:Parameswaran3, ref:Callin}. See also Gaherghetta et al \cite{ref:Gaherghetta, ref:Gaherghetta2}. \\
\indent We now proceed with the background line element, which is given by
\begin{align}
 & ds^2 = g_{\mu \nu } dx^{\mu } dx^{\nu } + a^2 ( d\theta ^2 + \sin^2{\theta } d\varphi ^2 )~,
 \label{eq:definition_metric} 
\end{align}
where $g_{\mu \nu } = \phi (\theta )\eta _{\mu \nu }$, $x^{\mu }$ are coordinates for our 4D spacetime, while $ 0\leq  \theta \leq \pi$ and $0\leq \varphi \leq 2\pi $  are coordinates for the extra 2D spherical surface with a constant radius $a$. \\
\indent  One of the most characteristic things is that the warp factor can be fixed almost completely from the positive energy condition of the 6D energy-momentum tensor (EMT) of bulk matter fields. One of possible results is given by
\begin{align}
 & \phi (\theta ) = \epsilon \exp{(a\sin^2{\theta })}~, 
 \label{eq:warp_facter}
\end{align}
where $4 \alpha \leq 1$ and $\Lambda a^2 \leq -2$. Here $\epsilon $ is an arbitrary constant and $\Lambda $ the 6D cosmological constant. In this approach we take a general EMT, which does not depend on a special choice of bulk matter fields.  \\
\indent We can easily see that the energy of any particle running along the geodesic line above is given by
\begin{align}
 & E = \Big( \frac{d\theta }{d\tau } \Big)^2 - \frac{A^2}{\phi ^2(\theta )}~,
 \label{eq:eq_of_motion}
\end{align}
where  $A$  is a constant. The potential has a maximum value $-A^2/(\epsilon e^{\alpha })^2$ at $\theta =\pi /2$ and the lowest value $-A^2/\epsilon ^2$  at $\theta =0, \pi $, so that any massive particle is rolling down into points $\theta =0, \pi $. As for massless particles we can see that they can extend into the 6D world.\\
\indent For the most general perturbation around the background $g_{IJ}$ we calculate the gravitational potential in our 4-dimensional universe. In this formulation we would like to point out that the traceless energy-momentum tensor is quite naturally appeared in our 6-dimensional model. Furthermore, we should stress that the complicated Einstein equations reduce to the simple differential equations for the spheroidal functions, which can be easily solved. \\
\indent The gravitational potential in our 4-dimensional universe is calculated to be composed of infinite series of massive Yukawa potentials coming from the KK mode, together with Newtonian's inverse law. The series of Yukawa type potentials can be summed up for the large n, say $n>n_0$ to give $\epsilon ^2/(a^2 r^3)$ near $r=0$. \\
\indent In Sec.\ref{sec:2} we give the warp factor and discuss gravitational wave equations in the 6D world. In Sec.\ref{sec:3} eigenfunctions of differential operator of the Einstein equation are obtained. In Sec.\ref{sec:4}the gravitational potential is calculated by means of the Green functional method. The final section is devoted to concluding remarks. We prepare the Appendix A for fixing the warp factor from the requirement of the positive energy condition for the 6D EMT. 
%  
%  
%%%%%%%%%%%%%%%%%%%%%%%%%%%%%%%%%%%%%%%%%%%%%%%%%%
\section{Gravitational wave equation}\label{sec:2}
%%%%%%%%%%%%%%%%%%%%%%%%%%%%%%%%%%%%%%%%%%%%%%%%%%
%
%
The Einstein equation with the 6D metric $g_{IJ}$ is given by 
\begin{align}
 &  R_{IJ} - \frac{1}{2}g_{IJ} R + \Lambda g_{IJ} = \kappa T_{IJ}~,
  \label{eq:Einstein_eq}
\end{align}
where $\kappa = 8\pi G_6$, $R_{IJ} = R^{K}_{\ IKJ}$ and we follow notations in Wald's book \cite{ref:Wald}. This equation can be rewritten as
\begin{align}
 &  R_{IJ} - \frac{1}{2} g_{IJ} \Lambda = \kappa \Big( T_{IJ} - \frac{1}{4} g_{IJ} T \Big)~,
  \label{eq:Einstein1} 
\end{align}
with $T = g^{IJ} T_{IJ}$.  The factor $1/4$ is characteristic in the 6D model. \\
\indent  We look for background solutions of Eq. (\ref{eq:Einstein1}) for the warp factor $\phi (\theta )$ with the ansatz for EMT of bulk matter fields \cite{ref:Gogberashvili, ref:Agnilar, ref:Kanti, ref:Gogberashvili2, ref:Oda, ref:Kokado} 
\begin{align}
 & T_{I J} = - g_{I J} f_n(\theta )~,
 \label{eq:ansatz}
\end{align}
where
\begin{align}
 & T_{\mu \nu } = - g_{\mu \nu } f_1(\theta )~,
  \label{eq:Energy_momentum_tensor} \\
 & T_{55} = - g_{55} f_2(\theta )~,
  \nonumber \\
 & T_{66} = - g_{66} f_3(\theta )~.
  \nonumber
\end{align}
All other elements vanish. Relevant quantities with the metric (\ref{eq:warp_facter}) are inserted into the ($\mu \nu $) component of Eq. (\ref{eq:Einstein1}) to give

\begin{align}
 & \frac{ 3 \phi'^2 + \phi \phi'' + \cot{\theta } \phi \phi' }{\phi ^2}   + \frac{1}{2} \Lambda a^2 2 = - \frac{1}{4}\kappa a^2 ( f_2 + f_3)~.
 \label{eq:equation_mu_nu_0}
\end{align}
In the same way, we get 
\begin{align}
 & 4\frac{\phi'' }{\phi } - 1 + \frac{1}{2} \Lambda a^2 = - \frac{1}{4}\kappa a^2  (4f_1- 3f_2 + f_3)~,
 \label{eq:equation_55_0}
\end{align}
\begin{align}
 & 4\cot{\theta }\frac{\phi'}{\phi } - 1 + \frac{1}{2} \Lambda a^2 = - \frac{1}{4}\kappa a^2  (4f_1 + f_2 - 3f_3)~,
 \label{eq:equation_66_0}
\end{align}
for the (55) and (66) component respectively. From Eqs. (\ref{eq:equation_mu_nu_0})-(\ref{eq:equation_66_0}) we obtain
\begin{align}
 & f_1= - \frac{1}{\kappa  a^2} \Big[ 3\frac{ \phi'^2 + \phi \phi'' + \cot{\theta } \phi \phi' }{\phi ^2} -1 + \Lambda a^2 \Big]~,
 \label{eq:value_f1} \\
 &  f_2 = - \frac{1}{\kappa  a^2} \Big[ \frac{ 6 \phi'^2 + 4 \cot{\theta } \phi \phi' }{\phi ^2} + \Lambda a^2 \Big] ~,
 \nonumber \\
  & f_3 = - \frac{1}{\kappa  a^2} \Big[ \frac{ 6 \phi'^2 + 4\phi \phi'' }{\phi ^2}  + \Lambda a^2 \Big] ~.
  \nonumber
\end{align}
The left-hand side quantities $f_i$'s correspond to EMT $T_{IJ}$'s, whereas the right-hand sides are those of Ricci tensors $R_{IJ}$'s in Eq.(\ref{eq:Einstein_eq}). Note that the Ricci tensor $R_{IJ}$ is always divergence free, i.e., $\triangle _I R^{IJ}=0$ because of the Bianchi identity, where $\triangle _I$ is the covariant derivative. Hence also so is the EMT, i.e., $\triangle _I T^{IJ}=0$ for any function $\phi $. \\
\indent Now we would like to fix the functional form of $\phi $. We use the positive energy condition of the 6D EMT , in a region $0\leq \theta \leq \pi $, that is,
\begin{align}
 & u_{I} T^{IJ} u_{J} = f_1 + u_5 u^5 (f_1 - f_2) + u_6 u^6 (f_1 - f_3) \geq 0~,
 \label{eq:posivity_condeiton_1}
\end{align}
where $u_I$ is a unit time-like vector in 6-dimensions. We assume that the warp factor is given by $\phi = \epsilon \exp{(\alpha \sin^2{\theta })}$, where $\epsilon $ and $\alpha $ are arbitrary parameters. In order to fix them, $\phi $ is substituted into inequalities, $f_1\geq 0$, $f_1-f_2\geq 0$, and $f_1-f_3\geq 0$. Then we have
\begin{align}
 & \phi (\theta ) = \epsilon e^{(\alpha \sin^2{\theta })}~,
 \label{eq:def_fai}
\end{align}
with $4\alpha \leq 1$ and $\Lambda a^2 \leq -2$ [see Appendix A]. If Eq.(\ref{eq:def_fai}) is substituted into brackets in right-handed of Eq.(\ref{eq:value_f1}), we get our EMT, i.e., $f_n, n=1,2,3$,
\begin{align}
 & \kappa a^2 f_1 = 24 \alpha ^2 \sin^4{\theta } + \alpha (18 - 24 \alpha ) \sin^2{\theta }  + 1- \Lambda a^2  - 12\alpha ~,
  \label{eq:cal_f_1} \\
 & \kappa a^2 f_2 = 24 \alpha ^2 \sin^4{\theta } + \alpha (8 - 24 \alpha ) \sin^2{\theta }  - \Lambda a^2 - 8\alpha ~,
  \nonumber \\
 & \kappa a^2 f_3 = 40 \alpha ^2 \sin^4{\theta } +  \alpha (16 - 40 \alpha ) \sin^2{\theta } - \Lambda a^2 - 8\alpha ~,
 \nonumber
\end{align}
In Appendix B we give an example that the above EMT can be actually derived from usual field theories. \\
\indent Let us now consider the most general perturbation $g_{IJ}^{(1)} = h_{IJ}$ around the background metric $g_{IJ}^{(0)}$. We follow the technique previously obtained \cite{{ref:Kokado2}}. The line element is 
\begin{align}
 & ds^2 = g_{IJ} dx^I dx^J = (g_{IJ}^{(0)} + g_{IJ}^{(1)} )dx^{I} dx^{J} ~,
 \label{eq:definition_metric_a}
\end{align}
where
\begin{align}
 & g_{\mu \nu } = g^{(0)}_{\mu \nu }+g^{(1)}_{\mu \nu } = \phi ^2(\theta )\eta _{\mu \nu } + h_{\mu \nu } ~, \quad g_{\mu 5} = g^{(1)}_{\mu 5} = h_{\mu  5}, \quad g_{\mu 6} = g^{(1)}_{\mu 6}= h_{\mu 6}~, 
 \label{eq:metric_add_h} \\
 & g_{55} = g^{(0)}_{5 5}+g^{(1)}_{5 5 } = a^2 + h_{55}, \quad g_{66} = g^{(0)}_{6 6}+g^{(1)}_{6 6}=a^2 \sin^2{\theta } + h_{66}, \quad g_{56} = g_{56}^{(1)} = h_{56}~.
 \nonumber 
\end{align}
We put here the gauge conditions
\begin{align}
 & \partial _{K } h^{K}_{\ \mu } = 0~, \quad \partial _{K } h^{K}_{\ 5} = 0, \quad h_{5 6} =0~.
 \label{eq:gauge_condition}
\end{align}
\indent We now expand $R_{IJ}$, $T_{IJ}$ in order $h_{IJ}$ as
\begin{align}
 & R_{IJ} = R^{(0)}_{IJ} + R^{(1)}_{IJ}~, \quad T_{IJ} = T^{(0)}_{IJ} + T^{(1)}_{IJ}~.
 \label{eq:expand_h}
\end{align}
where
\begin{align}
 & T_{\mu \nu }^{(0)} = - g_{\mu \nu }^{(0)} f_1 (\theta ) = - \phi ^2 (\theta ) \eta _{\mu \nu } f_1(\theta )~,
 \label{eq:def_em_tensor_mn} \\
 & T_{55}^{(0)} = - g_{5 5}^{(0)} f_2 (\theta ) = - a^2 f_2 (\theta )~,
 \nonumber \\
 & T_{66}^{(0)} = - g_{6 6}^{(0)} f_3 (\theta ) = - a^2 \sin^2{\theta } f_3 (\theta )~.
 \nonumber
\end{align}
We shall restrict our attention to the most interesting case of a static particle with mass $m_0$ located at $\vec{x} = \vec{0}, \theta = 0$. In this case EMT is given by
\begin{align}
 & T_{I J}^{(1)} = \tau _{I J} (x)\delta (\theta) = m_0 \delta _{I} ^{0} \delta _{J} ^{0} \delta ^{(3)}(\vec{x}) \delta(\theta )~,
 \label{eq:em_tensor_static_mass} 
\end{align} 
$\tau _{I J}$ being in order $h_{I J}$. However, a more general form of $T_{I J}^{(1)}$ comes from by the definition $T_{I J} = - g_{I J} f_{n}$, i, e., the first order change of $T_{I J} \equiv \delta T_{I J} = -h_{I J} f_{n}(\theta ) -g_{I J}^{(0)} \delta f_n(\theta )$ with\footnote{According to our model in Appendix B, these parameters are explicitly given by
\begin{align}
 & c_1 = \frac{1}{2}\big( \partial ^{\theta } \psi \big)^2  + \frac{1}{2} F^5_{\ 6} F^{56}~,  \quad d_1 = \frac{1}{2} F_5^{\ 6} F^{56} + \frac{\mu ^2}{2} \big( A^6 \big)^2~,
 \nonumber \\
 & c_2 = - \frac{1}{2}\big( \partial ^{\theta } \psi \big)^2  - \frac{1}{2} F^5_{\ 6} F^{56}~,  \quad d_2 = - \frac{1}{2} F_5^{\ 6} F^{56} + \frac{\mu ^2}{2} \big( A^6 \big)^2~, 
 \nonumber \\
 & c_3 = \frac{1}{2}\big( \partial ^{\theta } \psi \big)^2  - \frac{1}{2} F^5_{\ 6} F^{56}~,  \quad d_2 = - \frac{1}{2} F_5^{\ 6} F^{56} - \frac{\mu ^2}{2} \big( A^6 \big)^2~.
 \nonumber
\end{align} }
\begin{align}
 & \delta f_{n}(\theta ) = \frac{df_n(\theta )}{dg^{(0)}_{K L}} h_{K L} = c_n h_{55} + d_{n} h_{66}~.
 \nonumber
\end{align}
Henceforth this form will be added to Eq.(\ref{eq:em_tensor_static_mass}), that is,
\begin{align}
 & T_{I J} = T_{I J}^{(0)} +  T_{I J}^{(1)} + \delta T_{I J}~.
 \label{eq:def_T_ab_1} 
\end{align}
\indent The first-order equation in Eq. (\ref{eq:Einstein1}) is given by
\begin{align} 
 & R_{IJ}^{\ (1)} - \frac{1}{2}g_{I J}^{\ (1)}\Lambda + \frac{1}{4}\kappa  g_{IJ}^{\ (1)}T^{(0)} + \frac{1}{4} \kappa g_{IJ}^{\ (0)} \tilde {T}^{(1)} - \kappa \delta T_{I J} +\frac{\kappa }{4} g_{I J}^{(0)} \delta T^{K}_{\ K}
 \label{eq:Einstein_Eq5_0} \\
 & = \kappa  \Big[T_{IJ}^{\ (1)}-\frac{1}{4} g_{IJ}^{\ (0)} T^{(1)}\Big] \equiv \kappa  \Sigma _{IJ}~.
 \nonumber
\end{align}
where $T = g^{IJ} T_{IJ} = T^{(0)} + T^{(1)} + \tilde{T}^{(1)} + \delta T^{K}_{\ K}$ with
\begin{align}
 & T^{(0)} = g^{(0)I J} T_{I J}^{(0)} = -4 f_1(\theta) - f_2(\theta) - f_3(\theta)~,
 \nonumber \\
 & T^{(1)} = g^{(0)I J} T_{I J}^{(1)} = \tau ^{\lambda }_{\ \lambda }(x) \delta (\theta )~,
 \nonumber \\
 & \tilde{T}^{(1)}= g^{(1)I J} T_{I J}^{(0)} = - h^{\lambda }_{\ \lambda } f_1(\theta) - h^{5}_{\ 5} f_2(\theta ) - h^{6}_{\ 6} f_3(\theta )~.
 \nonumber \\
 & \delta T^{K}_{\ K} = - h^{\lambda }_{\ \lambda  } f_1(\theta ) - h^{5}_{\ 5}f_2(\theta ) - h^{6}_{\ 6} f_3(\theta ) 
 \nonumber \\
 &  - 4 (c_1 h_{5 5} + d_1 h_{6 6} ) - (c_2 h_{5 5} + d_2 h_{6 6} )-(c_3 h_{5 5} + d_3 h_{6 6} )~.
\end{align}
The source terms $\Sigma _{IJ}$ are explicitly given by
\begin{align}
 & \Sigma _{\mu \nu }(x, \theta) = \Big[ \tau _{\mu \nu }(x) - \frac{1}{4} g_{\mu \nu }^{(0)}\tau (x) \Big]\delta (\theta) = \big( \delta _{\mu }^{\ 0}  \delta _{\nu }^{\ 0} + \frac{1}{4} \eta _{\mu \nu } \big) m_0 \delta^{(3)}(\vec{x}) \delta (\theta)~,
 \label{eq:source_munu} \\
 & \Sigma _{5 5}(x, \theta) = - \frac{1}{4} a^2 \tau (x) \delta (\theta) = \frac{1}{4}\Big(\frac{a}{\epsilon }\Big)^2 m_0\delta^{(3)}(\vec{x}) \delta (\theta)~,
 \nonumber \\
 & \Sigma _{6 6}(x, \theta) = - \frac{1}{4} a^2 \sin^2{\theta} \tau (x) \delta (\theta) = 0~,
 \nonumber 
\end{align}  
where $\tau (x)\equiv \tau ^{\lambda }_{\ \lambda }(x)$. 
We solve Eq. (\ref{eq:Einstein_Eq5_0}) under such approximation that the source term $\Sigma _{5 5}$ is negligible compared with $\Sigma _{\mu \nu }$, that is, $|\Sigma _{5 5}| \ll |\Sigma _{\mu \nu }|$. This will be realized by the assumption $(a/\epsilon ) \ll 1$. We can then put as $\Sigma _{5 5} \simeq 0$. \\
\indent In a previous work \cite{{ref:Kokado2}} we found special solutions of Eq.(\ref{eq:Einstein_Eq5_0}):
\begin{align}
 & h^{\lambda }_{\ \lambda } = h^{5}_{\ 5} = h^{6}_{\ 6} = 0~,
 \label{eq:special_solutions_munu}
\end{align}
\begin{align}
 & \partial _K h^{K}_{\ 6} = 0~,
 \label{eq:solution_56}
\end{align}
\begin{align}
 & h_{5 \mu } = h_{6 \mu} = 0~.
 \label{eq:solution_tmu_6mu}
\end{align}
Thanks to these results (\ref{eq:special_solutions_munu}), (\ref{eq:solution_56}) and (\ref{eq:solution_tmu_6mu}) togather with gauge conditions (\ref{eq:gauge_condition}), the ($\mu \nu $) components of Eq. (\ref{eq:Einstein_Eq5_0}) reduces to
\begin{align}
 & - \frac{1}{2\phi ^2} \Box h_{\mu \nu}  - \frac{1}{2 a^2 \sin^2{\theta }}\partial _\varphi^{\ 2} h_{\mu \nu} 
 \label{eq:munu_Einstein_eq_A} \\
 & -\frac{1}{2 a^2}\Big[\partial _\theta ^{\ 2} h_{\mu \nu } + \frac{\cos{\theta }}{\sin{\theta }}\partial _{\theta }h_{\mu \nu } - 2 \big( \frac{\phi ''}{\phi } +  \frac{{\phi '}^2}{\phi ^2} + \frac{\cos{\theta }}{\sin{\theta }}\frac{\phi '}{\phi }\big) h_{\mu \nu }  \Big] 
 \nonumber \\
 & = \kappa [ \tau _{\mu \nu }(x) - \frac{1}{4} g_{\mu \nu }^{(0)} \tau (x)]\delta (\theta )~,
 \nonumber
\end{align}
where $\Box = \eta ^{\mu \nu }\partial _\mu \partial _\nu $. \\
\indent Eq.(\ref{eq:munu_Einstein_eq_A}) can be solved by means of the Green function, which is given by eigenfunctions of the differential operator $L$ in the left-hand side. Separating the four-dimensional mass term by
\begin{align}
 & h_{\mu \nu }(x, \theta , \varphi ) = h_{\mu \nu }(0) \exp{(i k_{\mu }x^{\mu })}  f(\theta ) k(\varphi )~, \quad k_{\mu } k^{\mu } = -M^2~,
 \label{eq:mass_condition}
\end{align}
with  $h^{\lambda }_{\ \lambda }(0)=0$, and substituting $\phi =\epsilon \exp{(\alpha \sin^2 \theta )}$ into Eq. (\ref{eq:munu_Einstein_eq_A}), we get, for $0 \leq \theta \leq  \pi $,
\begin{align}
 & -\frac{1}{2a^2} \Big[ \partial _{\theta }^{\ 2}  + \frac{\cos{\theta }}{\sin{\theta }}\partial _{\theta }+Q(\theta ) + \frac{1}{\sin^2{\theta }}\partial _{\varphi }^2 \Big]h_{\mu \nu } \equiv L h_{\mu \nu }= 0~.
 \label{eq:h_equation}
\end{align}
where
\begin{align}
 & Q(\theta ) = \lambda  - c^2 z^2~, \quad z=\cos{\theta }~,
 \label{eq:def_Q} \\
 & \lambda = 4 \alpha  + \frac{M^2a^2}{\epsilon ^2}  (1-2\alpha )~, 
 \label{eq:def_lambda} \\
 & c^2 = 2\alpha (6 - \frac{M^2a^2}{\epsilon ^2})~,
 \label{eq:def_c}
\end{align}
for a small $\alpha $. Here we have used an approximation that $\phi = \epsilon \exp{(\alpha \sin^2{\theta } )}\simeq \epsilon (1+\alpha \sin^2{\theta })$. Eq. (\ref{eq:h_equation}) becomes a simply separable equation. Hence we get $(\partial _\varphi ^2 + m^2 )k(\varphi )=0$, where $m$ should take integral values, because $k(\varphi )$ is a $2\pi $-periodic function. Putting the boundary condition $k(0)=0$, we get $k(\phi )=A\sin(m\varphi )$. The equation for $f(\theta )$ is, therefore, given by 
\begin{align}
 & \Big[ \partial _{\theta }^{\ 2}  + \frac{\cos{\theta }}{\sin{\theta }}\partial _{\theta } + Q_1(\theta ) \Big] f(\theta ) = 0~.
 \label{eq:h_equation2}
\end{align}
where
\begin{align}
 & Q_1(\theta ) = \lambda  - c^2 z^2 - \frac{m^2}{1-z^2}~, 
 \label{eq:def_Q_1} 
\end{align}
%
%
%
%
%%================================================
\section{Eingenfunctions} \label{sec:3}
%%================================================
%
The equation (\ref{eq:h_equation2}) is known as the spheroidal wave equation \cite{ref:Abramowitz}. The regular solution is given by in terms of associated Legendre functions apart from  normalization factors \cite{ref:Abramowitz}
\begin{align}
 & f_{mn}(c, z) = \sum_{l} d_{l}^{mn} P^{m}_{m+l}(z)~,
 \label{eq:spheroidal_wave_eq}
\end{align}
with eigenvalues
\begin{align}
 & \lambda _{mn} = \sum^{\infty }_{k=0} l_{2k} c^{2k}~,
 \label{eq:eigenvalues}
\end{align}
where
\begin{align}
 & l_0 = n(n+1)~,
 \label{eq:value_l_0} \\
 & l_2 = \frac{1}{2} \big[ 1 -\frac{(2m-1)(2m+1)}{(2n-1)(2n+3)} \big]~.
 \label{eq:value_l_2}
\end{align}
Since $P_{m+l}^m (1)=0$  for  $m \neq 0$, we have $f_{mn} (z=1)=0$. In a calculation of the gravitational potential, we do not interest the $m \neq 0$ case, because the case has no contributions to the potential. So, in the following we set  m=0 in Eqs. (\ref{eq:spheroidal_wave_eq}), (\ref{eq:eigenvalues}) and (\ref{eq:value_l_2}). \\
\indent From the equation
\begin{align}
 & \lambda _{n} = l_0 + l_2 c^2 + l_4 c^4 + \cdots = 4 \alpha + \frac{M^2a^2}{\epsilon ^2} (1-2\alpha )
 \label{eq:lambda_n}
\end{align}
we have a mass formula
\begin{align}
 & \frac{M^2a^2}{\epsilon ^2} \simeq \frac{l_0 + 4\alpha (3 l_2 - 1)}{1 - 2\alpha (1-l_2)}~.
 \label{eq:mass_formula}
\end{align}
The $0$-mass is obtained when the denominator is zero. The approximate formula for $\alpha \simeq 0$ is
\begin{align}
 & \frac{M^2a^2}{\epsilon ^2} \simeq l_0 = n(n+1) ~.
 \label{eq:mass_formula_alpha_0}
\end{align}
These masses correspond to KK masses.\\
\indent Normalization of Eq. (\ref{eq:spheroidal_wave_eq}) with $m=0$ is given by \cite{ref:Abramowitz}
\begin{align}
 & \int^{1}_{-1} dz f^2_{n}(c, z) = \frac{2}{2n+1}~.
 \label{eq:Normalization}
\end{align}
Hence the normalization factor is given by $N=\sqrt{(2n+1)/2}$. \\
\indent To sum up, the mass-eigenfunctions of Eq.(\ref{eq:spheroidal_wave_eq}) are defined by
\begin{align}
 & h_n(z) = \sqrt{\frac{2n+1}{2}} f_{n}(c, z) = \sqrt{\frac{2n+1}{2}} \sum_l d_{nl} P_{l}(z)
 \label{eq:cal_h}
\end{align}
Finite mass eigenvalues are approximately given by
\begin{align}
 & M_n = \frac{\epsilon \sqrt{n(n+1) }}{a}
 \label{eq:mass_formula2}
\end{align}
From the formula $\sum_l d_{nl} =1$ \cite{ref:Abramowitz}, we see $h_n (c,z=1)=\sqrt{(2n+1)/2}$.
%
%
%%===============================================
\section{Green function} \label{sec:4} 
%%================================================
%
The Einstein equation (\ref{eq:h_equation}) for $h_{\mu \nu }$, neglecting $\varphi $-dependence, is given by 
\begin{align}
 & L h_{\mu \nu } = \kappa \Sigma_{\mu \nu } (x, \theta )~,
 \label{eq:f1_4}
\end{align}
where
\begin{align}
 & L = -\frac{1}{2\phi ^2(\theta )}\partial _{\lambda } \partial^{\lambda } - \frac{1}{2a^2}\big[ \partial_{\theta }^2 + \frac{\cos{\theta }}{\sin{\theta }}\partial _{\theta } + Q(\theta ) \big]~,
 \label{eq:Laplacian_4}
\end{align}
\begin{align}
 & Q(\theta ) = (\lambda - c^2 \cos^2{\theta })|_{M=0} ~,  
 \label{eq:cal_Q_z_x}
\end{align}
\begin{align}
 & \Sigma _{\mu \nu }(x, \theta ) = \big( \tau _{\mu \nu }(x) - \frac{1}{4} g^{(0)}_{\mu \nu } \tau (x)\big) \delta (\theta )~.
 \label{eq:Laplacian_4a}
\end{align}
Here $\Sigma_{\mu \nu } (x, \theta )$  is the traceless energy-momentum tensor, consistent with the traceless condition $h^{\mu }_{\mu }=0$.  \\ 
\indent The solution $h_{\mu \nu }$ can be derived by means of the Green function as
\begin{align}
 & h_{\mu \nu }(X) = \int d^5 X'~G_R(X, X') \kappa \Sigma_{\mu \nu } (X'), \quad X=(x, \theta )~,
 \label{eq:h_Green_fun}
\end{align}
The Green function is defined by
\begin{align}
 & L G_R(X, X') = \delta ^5 (X - X')~,
 \label{eq:def_Green_fun}
\end{align}
where
\begin{align}
 & G_R(X, X') = L^{-1} \sum_n h_n(X) h_n^{\dagger}(X')
 \label{eq:Cal_Green1} \\
% & = \sum_m \int \frac{d^4 p}{(2\pi )^4}\frac{1}{\frac{-1}{2\phi ^2{\theta }}(\partial _\lambda \partial ^\lambda - m^2)} e^{i p (x -x')} h_m (\theta ) h_m^{\dagger}(\theta ')
% \nonumber \\
 & = (-2) \sum_n \int \frac{d^4 p}{(2\pi )^4} \frac{e^{i p(x-x')}}{\frac{1}{\phi ^2(\theta )}(-p^2-M_n^2)}h_n(\theta ) h_n^{\dagger}(\theta ')~.
 \nonumber
\end{align}
Here $h_n(\theta )$  are mass eigenfunctions of the differential operator $L$. \\
\indent Following Tanaka et al. \cite{ref:Garriga} we define the stationary Green function by 
\begin{align}
 & G_R(\vec{x}, \theta ;\vec{x'}, \theta ') = \int _{-\infty }^{\infty } dt'  G_R(X, X')
 \label{eq:Cal_Green2} \\
 & = 2 \sum_n \int \frac{d^3 p}{(2\pi )^3} \frac{e^{i \vec{p}(\vec{x}-\vec{x}')}}{\frac{1}{\phi ^2(\theta )}(\vec{p}^2+M_n^2)}h_n(\theta ) h_n^{\dagger}(\theta ')
 \nonumber \\
% & = 2 \phi ^2(\theta ) \sum_m \frac{1}{4\pi r}e^{-mr}h_m(\theta ) h_m^{\dagger}(\theta ')
% \nonumber \\
 & = 2 \phi ^2(\theta ) \frac{1}{4\pi r}\Big[ h_0(\theta ) h_0^{\dagger}(\theta ) + \sum_{n \geq 1} e^{-M_{n}r}h_n(\theta ) h_n^{\dagger}(\theta ') \Big]~,
 \nonumber
\end{align}
where $r = |\vec{x} - \vec{x'}|$.  The gravitational potential between masses $m_0$ and $m_1$   separated by $r$ is given, by putting $\theta =\theta '= \vec{x'} =0$, as follows:
\begin{align}
 & V(r) = - \frac{1}{2} m_{1} h_{00}(r) = -\frac{3}{8} m_0 m_1 G_R(\vec{x}, 0; \vec{0}, 0)
 \label{eq:Cal_Green_fun3} \\
 & = - 6\pi G_{6} m_0 m_1 \phi ^2(0) \frac{1}{4\pi r} \Big[ h_0(0) h_0^{\dagger}(0) + \sum_{n \geq 1} e^{-M{n} r} h_n(0) h_n^{\dagger}(0) \Big]~,
 \nonumber 
\end{align}
According to Eqs. (\ref{eq:mass_formula_alpha_0}) and (\ref{eq:Normalization}), we have
\begin{align}
 & V(r) = - G_N \frac{m_0 m_1}{r} \Big( 1 + \sum_{n=1}^{\infty } \alpha _{n} e^{-M_{n} r} \Big)~,
 \label{eq:gravitational potential}
\end{align}
with
\begin{align}
 & G_N = \frac{15}{4} \epsilon ^{2} G_6~,
  \label{eq:defin_GN} \\
 & \alpha _{n} = \frac{2n+1}{5} , \quad \frac{M_n a}{\epsilon} \simeq \sqrt{n(n+1)} \quad (n\geq 1)~.
 \nonumber
\end{align}
%
%
%%===============================================
\section{Concluding remarks} 
%%================================================
%
\indent In the warped 6D world with an extra 2D sphere we have succeeded to calculate the gravitational potential, which is given by Eq. (\ref{eq:Cal_Green_fun3}) with (\ref{eq:gravitational potential}). We have taken a general EMT, which does not depend on a special choice of bulk matter fields. We have fixed the warp factor to be $\phi =\epsilon \exp{(\alpha \sin^2{\theta })}$ with $1/4\geq \alpha >0$ and $x=1-\Lambda a^2\geq 3$ from the positive energy condition of EMT. Actually we have made of an approximation that $\alpha $ is so small enough $1>>\alpha $. The 6D Einstein equation has reduced to the spheroidal differential equation, which can be easily solved.\\
\indent The infinite series of Yukawa type potentials coming from KK-modes can be summed up for the large n, say $n>n_0$. Namely, the gravitational potential reduces to
\begin{align}
 & V(r) = - G_N \frac{m_0 m_1}{r} \Big( 1 + \sum_{n=1}^{n_0 } \alpha _{n} e^{-M_{n} r} + f(r) \Big)~,
 \label{eq:gravitational potential2}
\end{align}
where
\begin{align}
 & f(r) = \frac{2}{5} x^{n_0} \Big[ \frac{n_0}{1-x} + \frac{1}{(1-x)^2}\Big]~,
 \nonumber
\end{align}
with $x = e^{-\epsilon r/a}$. The function $f(r)$ behaves as $\epsilon ^2/(a r)^2$ near $r=0$. Two latter correction terms to Newton's $1/r$ law is remarkable in the warped 6D world with the small extra 2D sphere, to be experimentally checked. \\
\indent The 4D Newton constant $G_N$ is given by $G_N=\epsilon ^2 G_6$. The smallness of $G_N$, therefore, may reflect that of $\epsilon $. If we choose $\epsilon =10^{-19}$ against $G_6=1$Ge$V^{-2}$, then we get the present value of $G_N=10^{-38}$ Ge$V^{-2}$. 
\section*{Acknowledgments}
We would like to express our deep gratitude to T. Okamura for many valuable discussions.
\appendix
%%%%%%%%%%%%%%%%%%%%%%%%%%%%%%%%%%%%%%%%%%%%%%%%%%
\section{The positive energy condition of EMT}\label{sec:appendA}
%%%%%%%%%%%%%%%%%%%%%%%%%%%%%%%%%%%%%%%%%%%%%%%%%%
%
%
\indent The warp factor $\phi (\theta )$ is expanded, around $\theta =0$, into the Tayler series, $\phi(\theta ) = \phi_0 +\phi_1\theta +\phi _2 \theta ^2 + \phi_3 \theta ^3+ \cdots$. This form is substituted into $f_i(\theta )$ to yield, dorpping the factor $1/(\kappa a^2)$.
\begin{align}
 & f_1 = - \frac{3\phi_1\phi_0 + \theta (12\phi_2 \phi_0 +6\phi_1^2 + (\Lambda a^2 - 1)\phi_0^2)}{\theta \phi(\theta ) ^2}\geq 0~,
 \label{eq:positive_condition_f1} \\
 & f_1 - f_2 = \frac{\phi_1\phi_0 + \theta (-4\phi_2 \phi_0 +4\phi_1^2 + \phi_0^2)}{\theta \phi(\theta ) ^2}\geq 0~,
 \label{eq:positive_condition_f1_f2} \\
  & f_1 - f_3 = \frac{-3\phi_1\phi_0 + \theta (-4\phi_2\phi_0  + \phi_0^2)}{\theta \phi(\theta ) ^2}\geq 0~,
 \label{eq:positive_condition_f1_f3}
\end{align}
In a limit $\theta \to 0$, we have consistent results $\phi_1 =0$,  $\phi_0/4\geq \phi_2 >0$,   $(1-\Lambda a^2)\phi_0 /12 \geq \phi _2 >0$. Hence we get $\phi (\theta )=\phi_0 + \phi_2 \theta ^2 + \cdots$, where $\phi_0/4 \geq \phi_2 > 0$,  $1-\Lambda a^2 \equiv x \geq 3$.

This form suggests us generally to put an ansatz for $\phi (\theta )$ 
\begin{align}
 & \phi (\theta ) = \epsilon \exp{(\alpha \sin^2{\theta })}~,
 \label{eq:define_phi_append}
\end{align}
where
\begin{align}
 & \frac{1}{4} \geq \alpha >0~, \mbox{ and } x\geq 3~,
 \label{eq:formura_ln}
\end{align}
This form is substituted into inequalities, $f_1\geq 0, f_1-f_2\geq 0$ and $f_1-f_3\geq 0$ to yield
\begin{align}
 & f_1 - f_3 = -16 \alpha ^2 s^4 + 2s^2 \alpha ( 1+ 8 \alpha ) + 1 - 4 \alpha ~.
 \label{eq:cal_f_1f_3}
\end{align}
Here $s=\sin{\theta }$. If $0\leq 4\alpha \leq 1$. this equation is always positive in a reason $0\leq s^2 \leq 1$. This can be seen by calculating roots of $f_1-f_3=0$ for $s^2$. The second equation yields 
\begin{align}
 &  f_1 - f_2 = 10 \alpha s^2 + 1 - 4 \alpha ~.
 \label{cal_f_1_f_3}
\end{align}
This is nonnegative when $0 \leq 4\alpha \leq 1$. Finally, $f_1$ is given by
\begin{align}
 & f_1 = 24 \alpha ^2 s^4 + s^2 \alpha (18 - 24 \alpha ) + x - 12\alpha ~,
  \label{eq:cal_f_1a} 
\end{align}
where $x= 1-\Lambda a^2$. % If $\alpha $ terms are neglected,Eq.(8)  is nonnegative for $x \geq 0$.  More generally if we set  $4\alpha =1- \delta $ with $\delta >0$, then we have $15-24 \alpha =15-6+6 \delta = 9+ 6 \delta > 0$. The last term becomes $x - 9\alpha = x - 9/4+9\delta /4 $. This term is positive because of $x\geq 3$. Hence we see that 
Eq. (\ref{eq:cal_f_1a}) is positive if $4\alpha \leq 1$ and $x \geq 3$ \\.
 \indent  In conclusion, the warp factor $\phi =\epsilon \exp{(\alpha \sin^2{\theta })}$ makes EMT positive when $4\alpha \leq 1$ and $x \geq 3$, (This means $\Lambda a^2 \leq -2$).
%
%%%%%%%%%%%%%%%%%%%%%%%%%%%%%%%%%%%%%%%%%%%%%%%%%%
\section{A model of EMT}\label{sec:appendB}
%%%%%%%%%%%%%%%%%%%%%%%%%%%%%%%%%%%%%%%%%%%%%%%%%%
%
We give an example that our EMT can be derived from usual field theories. We first start from EMT in a non-warped 6D space with $\phi =1$, then will arrive at the warped 6D space with $\phi \neq 1$.  Therefore, we consider here a scalar field $\psi $ and a vector field $A_a$ defined in the non-warped 6D space. The action is given by
\begin{align}
 & S = \int d^6 x \sqrt{-g} L = \int d^6 x \sqrt{-g}\Big[ -\frac{1}{2} \partial _a \psi \partial ^a \psi  + \psi ^2 - \frac{1}{4} F_{ab}F^{ab} - \frac{\mu ^2}{2} A_a A^a \Big]
 \label{eq:action_real_field} \\
 & = \int d^6 x a^2 \sin{\theta } \Big[ -\frac{1}{2}\partial _\theta  \psi \partial ^\theta  \psi  + \psi ^2  - \frac{1}{2} \partial _\theta A_6 \big( \partial ^\theta A^6 + 2 \frac{\cos{\theta }}{a^2 \sin{\theta }}A^6 \big) -\frac{\mu ^2}{2} A_6 A^6 \Big]~.
 \nonumber 
\end{align}
Here we have considered $\psi = \psi (\theta )$ and $A_a = (0, 0, 0, 0, 0, A_6(\theta ))$. Then we get equations of motion for $\psi $ and $A_6$.
\begin{align}
 & \partial _{\theta }^{2} \psi  + \frac{\cos{\theta }}{\sin{\theta }} \partial _{\theta } \psi + 2 \psi =0~,
 \label{eq:EOM_psi} \\
 & \partial _\theta ^{2} A^6  + 3 \frac{\cos{\theta }}{\sin{\theta }} \partial _\theta A^6 - (\mu ^2a^2 + 2) A^6 =0~,
 \label{eq:EOM_A6}
\end{align}
solutions are
\begin{align}
 & \psi = A \cos{\theta }~,
 \label{eq:solution_psi} \\
 & A^6 = B\big( 1 + \frac{M + 2}{8} \sin^2{\theta } \big), \quad M = \mu ^2 a^2~,
 \label{eq:solution_A6}
\end{align}
where we have neglected the order $\sin^4{\theta }$ in Eq.(\ref{eq:solution_A6}), because we are enough to consider near $\theta =0$ (see the introduction). \\
\indent According to the formula of EMT we get
\begin{align}
 & T_{ab} = - \frac{2}{\sqrt{-g}} \frac{\delta \big( \sqrt{-g} L \big)}{\delta g^{ab}}~,
  \label{eq:formula_T} \\
 & = \partial _a \psi \partial _b \psi + F_{ac} F_{b}^{\ c} + \mu ^2 A_{a} A_{b} + g_{ab} L~,
 \nonumber
\end{align}
to yield
\begin{align}
 & T_{\mu \nu } = - g_{\mu \nu } \big( \frac{1}{2} \partial _\theta \psi \partial ^{\theta }\psi -\psi ^2 + \frac{1}{2} F_{5 6} F^{5 6} + \frac{1}{2} \mu ^2 A_6 A^6 \big) = - g_{\mu \nu } f_1 ~,
 \label{eq:cal_EOT} \\
 & T_{5 5} = - g_{5 5} \big( - \frac{1}{2} \partial _\theta \psi \partial ^{\theta }\psi -\psi ^2 - \frac{1}{2} F_{5 6} F^{5 6} + \frac{1}{2} \mu ^2 A_6 A^6 \big) = - g_{5 5} f_2 ~,
 \nonumber \\
 & T_{6 6} = - g_{6 6} \big( \frac{1}{2} \partial _\theta \psi \partial ^{\theta }\psi -\psi ^2 - \frac{1}{2} F_{5 6} F^{5 6} - \frac{1}{2} \mu ^2 A_6 A^6 \big) = - g_{6 6} f_2 ~.
 \nonumber
\end{align}
Substituting solutions (\ref{eq:solution_psi}) and (\ref{eq:solution_A6}) into Eqs.(\ref{eq:cal_EOT}), we have
\begin{align}
 & f_1 = \frac{3}{2} A^2 \sin^2{\theta } - A^2 + \frac{3}{2} M^2 B^2 \sin^2{\theta } + 2 B^2~,
 \label{eq:value_f1B} \\
 & f_2 = \frac{1}{2} A^2 \sin^2{\theta } - A^2 - \frac{1}{2} M^2 B^2 \sin^2{\theta } - 2 B^2~,
 \nonumber \\
 & f_3 = \frac{3}{2} A^2 \sin^2{\theta } - A^2 - \frac{3}{2} M^2 B^2 \sin^2{\theta } - 2 B^2~,
 \nonumber
\end{align}
whereas $f_n'$s should be given by Eqs.(\ref{eq:cal_f_1}). The resulting three equations are, dropping the factor $\kappa a^2$, 
\begin{align}
 & f_1 = \alpha (18 - 24 \alpha ) \sin^2{\theta } + 1 - \Lambda a^2  - 12\alpha =  \frac{3}{2}\big( A^2  + M^2 B^2 \big) \sin^2{\theta } - A^2 + 2 B^2 ~,
 \label{eq:value_f1_B2} \\ 
 & f_2 = \alpha (8 - 24 \alpha ) \sin^2{\theta }  - \Lambda a^2 - 8\alpha =  \frac{1}{2}\big( A^2  - M^2 B^2 \big) \sin^2{\theta } - A^2 - 2 B^2 ~,
 \label{eq:value_f2_B2} \\
 & f_3 =  \alpha (16 - 40 \alpha ) \sin^2{\theta } - \Lambda a^2 - 8\alpha  = \frac{3}{2}\big( A^2  - M^2 B^2 \big) \sin^2{\theta } - A^2 - 2 B^2 ~,
 \label{eq:value_f3_B2}
\end{align}
where we have neglecting the order of $\sin^4{\theta }$ in the same reason above and $a^4$ has been absorbed into $B^2$.  Here constant terms should totally vanish.
\begin{align}
 &  1 - \Lambda a^2  -12 \alpha   = 2 B^2 - A^2~, \quad - \Lambda a^2 - 8\alpha = -2 B^2 - A^2 ~.
 \label{eq:solution_constant_term}
\end{align}
Hence we have
\begin{align}
 & B^2 = \frac{1}{4} - \alpha ~.
 \label{eq:solution_B_contant_part}
\end{align}
Equations (\ref{eq:value_f1_B2}), (\ref{eq:value_f2_B2}) and (\ref{eq:value_f3_B2}) then reduce to
\begin{align}
 & \alpha (18 - 24 \alpha ) =  \frac{3}{2}( A^2  + M^2 B^2 ) ~,
 \label{eq:equaiton_f1_B2_s2} \\ 
 & \alpha (8 - 24 \alpha ) = \frac{1}{2} ( A^2  - M^2 B^2 ) ~,
 \label{eq:equation_f2_B2_s2} \\
 & \alpha (16 - 40 \alpha ) = \frac{3}{2} ( A^2  - M^2 B^2 ) ~.
 \label{eq:equation_f3_B2_s2}
\end{align}
From Eqs.(\ref{eq:equation_f2_B2_s2}) and (\ref{eq:equation_f3_B2_s2}) we have $\alpha =1/4$. However, these equations are approximate equations. The true value may be given by $\alpha =1/4-\delta $, where $\delta \ll 1$. Hence we get solutions
\begin{align}
 & A^2 = \frac{3}{2} - 2 \delta~, \quad B^2 = \frac{1}{2 M} - \frac{8}{3 M}\delta ~.
 \label{eq:solution_A_b}
\end{align}
On the other hand, from Eq. (\ref{eq:solution_B_contant_part}) we have $B^2=\delta $, so that $\delta = 1/(2M)$, $M \gg 1$. \\
\indent This means that we have succeeded to construct our EMT from that of usual field theories. 
%
%\begin{thebibliography}{000} %for 3 digits
%\begin{thebibliography}{00}  %for 2 digits

\end{document}